\documentclass[a4paper,fleqn,usenatbib]{mnras}

\pdfoutput=1

\usepackage{newtxtext,newtxmath}

\usepackage[T1]{fontenc}
\usepackage{ae,aecompl}

\usepackage{graphicx}	
\usepackage{amsmath}	
\usepackage{pgfplots}

\newcommand{\comment}[1]{}

\title[GRMHD simulations of accreting, spinning black hole binaries]{Minidisk Influence on Flow Variability in Accreting Spinning Black Hole Binaries: Simulations in Full General Relativity}

\author[Bright \& Paschalidis]{
Jane C. Bright,$^{1}$
Vasileios Paschalidis,$^{1,2}$
\\
$^{1}$ Department of Astronomy, University of Arizona, Tucson, AZ 85726, USA\\
$^{2}$ Department of Physics, University of Arizona, Tucson, AZ 85726, USA\\
}

\pubyear{2022}

\begin{document}
\label{firstpage}
\pagerange{\pageref{firstpage}--\pageref{lastpage}}
\maketitle

\begin{abstract}
We perform magnetohydrodynamic simulations of accreting, equal-mass
binary black holes in full general relativity focusing on the
effect of spin and minidisks on the accretion rate and Poynting
luminosity variability. We report on the structure of the minidisks
and periodicities in the mass of the minidisks, mass accretion rates,
and Poynting luminosity. The accretion rate exhibits a quasi-periodic
behavior related to the orbital frequency of the binary in all systems
that we study, but the amplitude of this modulation is dependent on
the existence of persistent minidisks. In particular, systems that are
found to produce persistent minidisks have a much weaker modulation of
the mass accretion rate, indicating that minidisks can increase the
inflow time of matter onto the black holes, and dampen out the
quasi-periodic behavior. This finding has potential consequences for
binaries at greater separations where minidisks can be much larger and
may dampen out the periodicities significantly.

\end{abstract}

\begin{keywords}
black hole physics---gamma-ray burst: general---gravitation---gravitational waves---stars: neutron
\end{keywords}



\section{Introduction}

Supermassive black hole (SMBH) binaries are expected to form in gas
rich environments as a result of galaxy mergers
~\cite{Rodriguez:2009}. The central SMBHs from each galaxy in the
merger are driven to form a bound binary due to dynamical friction,
followed by stellar and other environmental interactions which can
bring them to close separations, where the subsequent evolution is
driven by gravitational wave emission (see, e.g.,
\cite{Mayer:2007}). SMBH binaries are particularly promising systems
for multi-messenger astronomy with gravitational waves. Gravitational
waves from SMBH binaries are expected to be detectable by Pulsar
Timing Arrays or by future space-based gravitational wave
observatories such as the Laser Interferometer Space Antenna (LISA)
~\cite{Hobbs:2010, LISA}, ~\cite{Amaro-Seoane:2022},
~\cite{Barausse:2020}, ~\cite{Arun:2022}.

Theoretical models make predictions about what the environments around
accreting SMBHBs could look like. When the gas around the binary has
sufficient angular momentum, the two black holes are expected to be
surrounded by a circumbinary accretion disk with the black holes
located in a low density central ``cavity'' that is cleared by the
binary tidal torques \cite{Artymowicz:1994}, \cite{Milosavljevi:2005},
\cite{Kocsis:2012}, \cite{MacFadyen:2008}. Accretion into the central
cavity then proceeds through two tidal streams, and can form minidisks
around the individual black holes. Shock heating of the minidisks may
make them responsible for enhancement of emission in the hardest parts
of the electromagnetic spectrum, see e.g., \cite{Sesana:2012},
\cite{Roedig:2014}, \cite{Farris:2015}. 

Binary systems within the gravitational-wave driven regime cannot be
observationally resolved by existing telescopes unless they are
extremely close to our Galaxy. Furthermore, it can be challenging to
distinguish a binary from a regular single black hole system
electromagnetically. Periodicities in quasar light curves may be able
to help identify systems containing binaries and distinguish them from
single black hole systems. Therefore, the key frontier in theoretical
work to-date has been to identify smoking-gun electromagnetic
signatures that accompany the gravitational wave signal and which can
be used to guide electromagnetic observations. Several mechanisms have
been proposed that may cause periodicities from binaries: Doppler
boosts along our line of sight of emission from gas bound to the
moving black holes \cite{D'Orazio:2015}, gravitational lensing of the
accretion onto one of the black holes by the companion black hole
\cite{D'Orazio:2018}, and variability in accretion rates onto the
black holes in the binary (see also \cite{Bogdanovic:2022} for a
recent review of electromagnetic emission from SMBH binary
mergers). Periodic accretion has been associated with an over-dense
``lump" feature forming on the inner edge of the circumbinary disk
caused by material from the accretion steams being flung outward and
breaking the axisymmetry. This lump then modulates the accretion onto
the minidisks, feeding more material to the black holes in a periodic
way \cite{Noble:2012}, \cite{Shi:2012}, \cite{Farris:2014},
\cite{Gold:2014a}, \cite{Noble:2021}. Recent studies suggest that the
periodicities in accretion rate are modulated by the lump relating to
the beat frequency of the lump's orbital frequency and the binary's
orbital frequency \cite{Bowen:2018}, \cite{Bowen:2019},
\cite{Combi:2022}.

However, variability of the mass accretion rate should depend on the matter
inflow time from the minidisks. In particular, if the inflow time from
the minidisks is longer than the modulation of the accretion flow due
to the binary orbit, then one could expect that such periodicities can
be phased out by the minidisks, in which case periodicities in the
electromagnetic lightcurves could not be attributed to accretion rate
modulation by the binary orbit. Interestingly many relativistic
studies of accreting binaries near merger do not exhibit large and
persistent minidisks, while Newtonian studies with parametrized inflow
time from the minidisks have reported such minidisks. The inflow time
from the minidisks is an unresolved question, and it could depend on
many physical parameters including black hole spin magnitude and
orientation, binary mass ratio, magnetic flux in the circumbinary
disk, et cetera. Given that periodicities in quasar lightcurves are
often viewed as modulations due to the binary orbit, it is important
to resolve when such quasi-periodic behavior in the accretion rate
onto the black holes is possible. This requires that we determine if
and when the inflow time from the minidisks becomes comparable or
longer than the binary orbital period especially for binaries in the
gravitationally wave driven and highly dynamical spacetime
regime. Another question is whether such binaries at relativistic
orbital separations are relevant at all for current electromagnetic
separations. In this work we begin to address these questions.

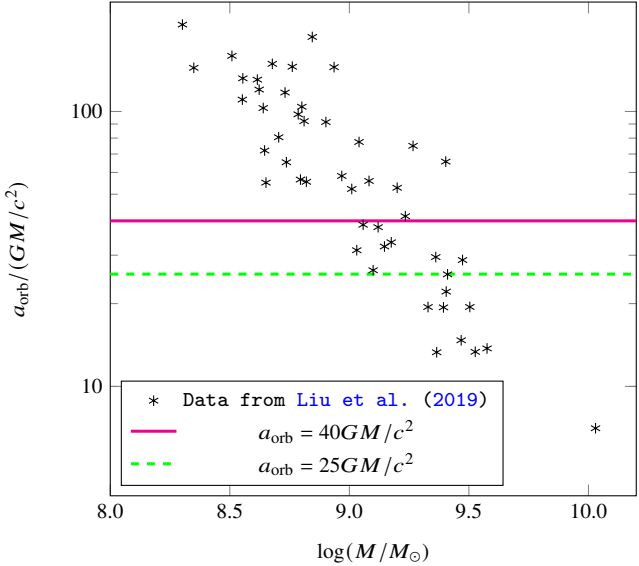
\begin{figure}
 \vspace{0.2cm}
\begin{center}  
\begin{tikzpicture}
  \begin{axis}[
    width=0.48\textwidth,height=3.2in,
    xmin = 8, xmax = 10.2,
    ymin = 4, ymax = 250,
    ymode = log,
    xlabel={$\log(M/M_\odot$)},
    ylabel={$a_{\rm orb}/(GM/c^2)$},
    legend entries={{\texttt{Data from \cite{Liu:2019gxe}}},{$a_{\rm orb}=40GM/c^2$},{$a_{\rm orb}=25GM/c^2$}},
    y tick label style={
      /pgf/number format/.cd,
      fixed,
      fixed zerofill,
      precision=2,
      /tikz/.cd
    },
    xtick={8,8.5,9,9.5,10},
    xticklabels={8.0,8.5,9.0,9.5,10.0},
    ytick={10,100},
    yticklabels={10,100},
    legend style={at={(0.02,0.02)}, anchor=south west},
    ]
    \addplot[only marks, mark=asterisk] table{
      7.92785   433.069
      7.99241   627.893
      8.3019   206.971
      8.34937   144.138
      8.50886   159.314
      8.55443   131.969
      8.55253   110.457
      8.61519   130.537
      8.62278   120.105
      8.63987   102.902
      8.64557   72.1001
      8.65127   55.0917
      8.67785   148.985
      8.70443   80.5638
      8.73101   117.174
      8.76139   145.478
      8.73671   65.3652
      8.78608   97.6096
      8.80127   104.005
      8.81076   92.2173
      8.79557   56.5313
      8.82025   55.4826
      8.84494   186.641
      8.93608   144.956
      8.9019   91.5171
      8.96835   58.2545
      9.01013   52.2418
      9.04051   77.4398
      9.03101   31.2781
      9.05759   38.7277
      9.08228   55.874
      9.09937   26.4499
      9.12025   37.9424
      9.14684   32.2668
      9.17532   33.4238
      9.2   52.7731
      9.23418   41.5821
      9.26646   74.9798
      9.32911   19.4401
      9.36139   29.5323
      9.36519   13.3069
      9.39367   19.3857
      9.40506   22.1211
      9.41076   25.5602
      9.40316   65.706
      9.47342   28.81
      9.46772   14.712
      9.5038   19.4492
      9.52658   13.3572
      9.57595   13.742
      10.0297   7.04281
    };
    \addplot[magenta,line width=0.4mm,domain=8:10.2]{ 40   };
    \addplot[green,line width=0.4mm,dashed,domain=8:10.2]{ 25.6   };
  \end{axis}
\end{tikzpicture}
\caption{
  BBH orbital separation ($a_{\rm orb}$ in
      units of $GM/c^2$) vs binary total mass $M$ for candidate
      SMBBHs. Asterisks are data from 
      \protect\cite{Liu:2019gxe}
      The magenta solid (green dashed) line designates an orbital separation of
      $40GM/c^2$ ($25GM/c^2$). Candidates with $a_{\rm orb}\lesssim
      25GM/c^2$ are in the highly dynamical spacetime
      regime. 
      \vspace{-0.1cm}} \label{sepvsM} \end{center}
\end{figure}

Photometric surveys have been implemented to perform systematic
searches for SMBHBs identifiable through periodicities in their light
curves, leading to about 200 current SMBH binary candidates
\cite{Graham:2015}, \cite{Charisi:2016}.  SMBHBs at relativistic
orbital separations of a few tens to a hundred gravitational radii are
likely observationally relevant. This is demonstrated explicitly in
Fig.~\ref{sepvsM}, where we plot the reported orbital separation
vs. mass for 50 candidate SMBBHs with 3 or more cycles of periodicity
in their lightcurves as compiled
in~\cite{Liu:2019gxe}. In~\cite{Liu:2019gxe} the orbital separation is
given in pc, so here we have converted the orbital separation to units
of gravitational radius using the mass estimate for these systems. As
is clear from the plot about 20 candidates are inferred to have
orbital separation $\lesssim 40 GM/c^2$ (where $M$ is the total binary
gravitational mass), and about 10 have an orbital separation $\lesssim
25GM/c^2$, where general relativity is particularly important to
determine the magnetohydrodynamic flow onto the SMBH
binaries. Therefore, binaries at relativistic orbital separations are
observationally relevant.

Circumbinary accretion has been investigated through both Newtonian
and relativistic simulations. Newtonian studies that have studied
circumbinary accretion include \cite{MacFadyen:2008},
\cite{D'Orazio:2013}, \cite{D'Orazio:2016}, \cite{Munoz:2016},
\cite{Miranda:2017}, \cite{Derdzinski:2019}, \cite{Munoz:2019},
\cite{Mosta:2019}, \cite{Duffell:2020}, \cite{Zrake:2021},
\cite{Munoz:2020}, .\cite{Munoz:2020b}, \cite{Derdzinski:2021}, and
Newtonian studies that investigate minidisks include
\cite{Farris:2014}, \cite{Farris:2015b}, \cite{Tang:2018},
\cite{Moody:2019}, \cite{TIede:2020}. Studies that use approximate
spacetime metrics include \cite{Bowen:2017}, \cite{Bowen:2018},
\cite{Bowen:2019}, \cite{Combi:2022}. Work that incorporates full GR
and MHD include \cite{Farris:2012}, \cite{Giacomazzo:2012},
\cite{Gold:2014a}, \cite{Gold:2014b}, \cite{Paschalidis:2021},
\cite{Cattorini:2021}, and see \cite{Gold:2019} for a recent
review of work on circumbinary accretion incorporating relativistic
effects. Newtonian studies are often conducted as 2D $\alpha$-disk
viscous hyrodynamics models which have the benefit of being able to
evolve for many orbits with less computational expense, but must
exclude the inner parts of the domain and impose ad hoc inflow
boundary conditions. Such boundary conditions should be informed by
relativistic calculations since the inflow time is a parameter in such
calculations. General relativity is paramount to treat these inner
regions self-consistently.

In our previous work, \cite{Paschalidis:2021}, we performed the first
fully general relativistic simulations of circumbinary accretion onto
spinning black holes and investigated the conditions under which
persistent minidisks form around the individual black holes. We found
that the accretion streams from the circumbinary disk form minidisks
whenever there are stable circular orbits around each black hole's
Hill sphere. This condition is met when the Hill sphere is
significantly larger than the effective innermost stable circular
orbit (ISCO). As black hole spin influences the radius of the ISCO, we
found that spin plays a crucial role in whether minidisks can form at
relativistic orbital separations. We found that at a separation of
$d=20GM/c^2$, black holes with a dimensionless spin parameter of
$\chi\equiv cJ/Gm^2=0.75$ (where $J$, $m$ are the black hole angular
momentum and mass, respectively) and $\chi=0$ were able to form
minidisks, while black holes with $\chi=-0.75$ were not. The latter is
because retrograde spin increases the size of the ISCO significantly,
thereby not allowing stable orbits within the Hill sphere at the
orbital separations probed by these simulations.

In this work, we extend the simulations presented in our previous work
\cite{Paschalidis:2021} and present a detailed analysis of the new
simulation data. In particular, we examine the structure of the
minidisks in greater detail, and investigate the effect that minidisks
and black hole spin have on the binary accretion and Poynting
outflows, and the quasi-periodic behaviors thereof. One of our most
important findings is that {\it persistent minidisks can dampen the
  strength of the modulation in the accretion rate}. While we do not
investigate in detail the reason why this occurs in this work, our
simulations suggest that the inflow time from the minidisks begins to
become comparable to the minidisk feeding timescale through the
circumbinary accretion streams. This is an important course of study
as it can have a significant impact on the observed electromagnetic
radiation from black hole binaries, and could affect the
interpretation of observed periodicities for systems with much larger
minidisks at larger binary orbital separations. This will be the topic
of study of future work.

We denote the individual black hole masses as $m$ (all our models
contain equal mass binaries $m_1=m_2=m$). We adopt geometrized units,
$G=c=1$, throughout the paper.

The rest of the paper is structured as follows: in Section
\ref{methods} we present our methods, including a description of the
models, and numerical methods we adopt in our evolutions. In Section
\ref{diagnostics} we report our diagnostics used to characterize the
accretion flow, EM signatures, minidisk structure, and Fourier
analysis. In Section \ref{results} we present the results from
analyzing our simulation data. In Section \ref{conclusions} we
summarize our findings and discuss future work.

\section{Methods}
\label{methods}

We employ the following set of assumptions/approximations in each of our models: 1)
the black holes are initially on quasi-circular orbits, 2) the
self-gravity of the disk is negligible in comparison to the gravity of
the binary, 3) the disk is well described by ideal MHD, 4) we do not
treat radiative feedback, heating or cooling.

\subsection{Initial Data}

For a detailed description of the magnetohydrodynamic and spacetime initial data used to start our simulations we refer the reader to~\cite{Paschalidis:2021}.
Here we only describe the different black hole configurations 
which were initially set on quasi-circular orbits at a coordinate
separation of 20M. We consider equal-mass black hole binaries in four
spin configurations: $\chi_1=\chi_2=0$ (nonspinning case labeled
$\chi_{00}$), $\chi_1=\chi_2=0.75$ (case $\chi_{++}$),
$\chi_1=\chi_2=-0.75$ (case $\chi_{--}$), $\chi_1=-\chi_2=0.75$ (case
$\chi_{+-}$). Where $\chi_1$, $\chi_2$ are the dimensionless spins of
each black hole, and $+$ ($-$) sign indicates spin aligned
(anti-aligned) with the orbital angular momentum.

\subsection{Evolution Equations and Methods}

We use the general relativistic magnetohydrodynamics (GRMHD)
adaptive-mesh-refinement (AMR), dynamical spacetime code
of~\cite{Etienne:2010}, \cite{Etienne:2012}, which employs the
Cactus/Carpet infrastructure \cite{Goodale:2003},
\cite{Schnetter:2004}. The code has been extensively tested and used
to study numerous systems involving compact objects and magnetic
fields.

We evolve the spacetime metric by solving Einstein's equations in the BSSN formulation \cite{Shibata:1995}, \cite{Baumgarte:1998}, coupled to the moving-puncture gauge conditions \cite{Baker:2006}, \cite{Campanelli:2006}, with the equation for the shift vector cast in first-order form, as in \cite{Hinder:2013}. The shift vector parameter $\eta$ is set to $\eta = 1.375/M.$

We evolve the matter and magnetic fields by solving the ideal GRMHD equations in flux-conservative form (see Eqs. 27-29 in \cite{Etienne:2010}) using a high-resolution shock capturing scheme. We enforce the zero-divergence constraint for the magnetic fields by solving the induction equation using a vector-potential formulation (see Eq. 9 in \cite{Etienne:2012}). For our EM gauge choice, we use the generalized Lorenz gauge condition developed in \cite{Farris:2012}, which avoids the development of spurious magnetic fields across the AMR levels, setting the generalized Lorenz gauge damping parameter to $\xi = 7/M$. 

\section{Diagnostics}
\label{diagnostics}
\subsection{MHD Flow Diagnostics}
We compute a set of diagnostics to help characterize the MHD flow and
the structure and influence of the minidisks. These diagnostics
include: 1) The accretion rate $\dot{M}$ as defined in
\cite{Farris:2010}, where we calculate both the total accretion rate
onto the binary and the accretion rate onto each individual black
hole. 2) The mass within the Hill spheres (i.e. mass of the minidisk
in cases where minidisks are present). The Hill Sphere radius is
calculated from the Newtonian formula $r_{Hill} = 0.5(q/3)^{1/3}d$,
where $q$ is the mass ratio and $d$ is the binary separation, and we
integrate the total rest mass within this radius centered on each
black hole. 3) The EM Poynting luminosity $L_{EM}$ on the surface of
coordinate spheres $S$ is computed as $L_{EM} = \oint_S
{T_{0}}^r_{,(EM)}dS$, where ${T_{\mu}}^\nu_{,(EM)}$ is the EM
stress-energy tensor.

\subsection{Minidisk Structure Diagnostics}

The minidisk structure parameters we calculate are centered on one of
the orbiting black holes, but all quantities are computed in the
binary center-of-mass frame, as opposed to boosting into the orbiting
black hole's frame. We expect that this would introduce corrections of
order $O(v^2)\sim 10\%$, but since the reported diagnostics below are not
gauge invariant we do not perform a boost.  We use the notation
$\varpi$ to indicate the cylindrical radius centered on the orbiting
black hole, rather than from the center of the grid.

The minidisk structure parameters we compute are 1) The surface
density profile $\Sigma (\varpi)$ of the minidisks, which is computed
as defined in \cite{Farris:2011} as $\Sigma = \frac{1}{2\pi}
\int^{2\pi}_0 \int_{z \geq 0} \rho_0 u^t \sqrt{-g} dz d\phi $.  2) The
scale height of the minidisks $H(\varpi)/\varpi$, computed as
$\Sigma/\rho_0(z=0)$.  3) The effective viscosity parameter
$\alpha(\varpi)$ of the minidisks which is calculated as the
approximate Shakura-Sunyaev stress parameter and computed as $\alpha =
\frac{ \langle T^{EM}_{r\phi}\rangle}{\langle P\rangle}$, where
$T^{EM}_{r\phi}$ is the orthonormal component of the Maxwell
stress-energy tensor evaluated using the tetrad in \cite{Penna:2010},
and $P$ the pressure. We stress again that none of these diagnostics
are gauge-invariant, but moving puncture coordinates are sufficiently
well-behaved that the reported quantities provide intuition into the
structure of these flows.

\subsection{Fourier Analysis}

\comment{
\vp{This section should be titled diagnostics, and it should report what diagnostic quantities and how they are computed. So, reference eqs. in other papers and move the equations you have further below here -- see my previous papers.}}

Periodicities are examined by performing Fourier analysis on the time
series of several of the output diagnostics. The time series data is
treated as follows: first, we subtract the running average
by smoothing the data, which is performed using a one-dimensional
Gaussian filter, 
and subtract this smoothed data from the un-smoothed data in order to remove the more
general underlying average behavior and isolate the periodic behavior. This
smooth-subtracted data is then windowed using the Tukey windowing
function. The data is then ``zero-padded", i.e., zeros are added to
the end of the data in order to produce a smoother function in
frequency space after the Fourier transform is performed without
affecting the underlying shape of the function. We then perform a
one-dimensional discrete Fourier transform using the Fast Fourier
Transform (FFT) algorithm as implemented in Python using the NumPy Real FFT.
The power spectral density (PSD) is then computed as
$\mid FFT \mid ^2$. We normalize the frequencies in the PSD to the
orbital frequency of the binary ($f_{orb})$, which is computed from
the average gravitational wave frequency of the $l=2, m=2$ mode.

\section{Results and discussion}
\label{results}

\subsection{Minidisk Structure}
\label{minidisk_structure}

\comment{
\vp{Somewhere it should be stated that these are centered on the orbiting black holes, but all quantities are as computed in the binary center-of-mass frame. We expect that compared to boosting into the frame of the orbiting black holes would bring corrections of order $O(v^2)\sim 10\%$.}}

Here we provide a discussion of the structure of the minidisks, which
may be used to inform 2D Newtonian simulations that prescribe minidisk
quantities, and especially when such studeis continue into the
relativistic regime. 

\begin{figure}
  \centering
  \includegraphics[width=0.47\textwidth]{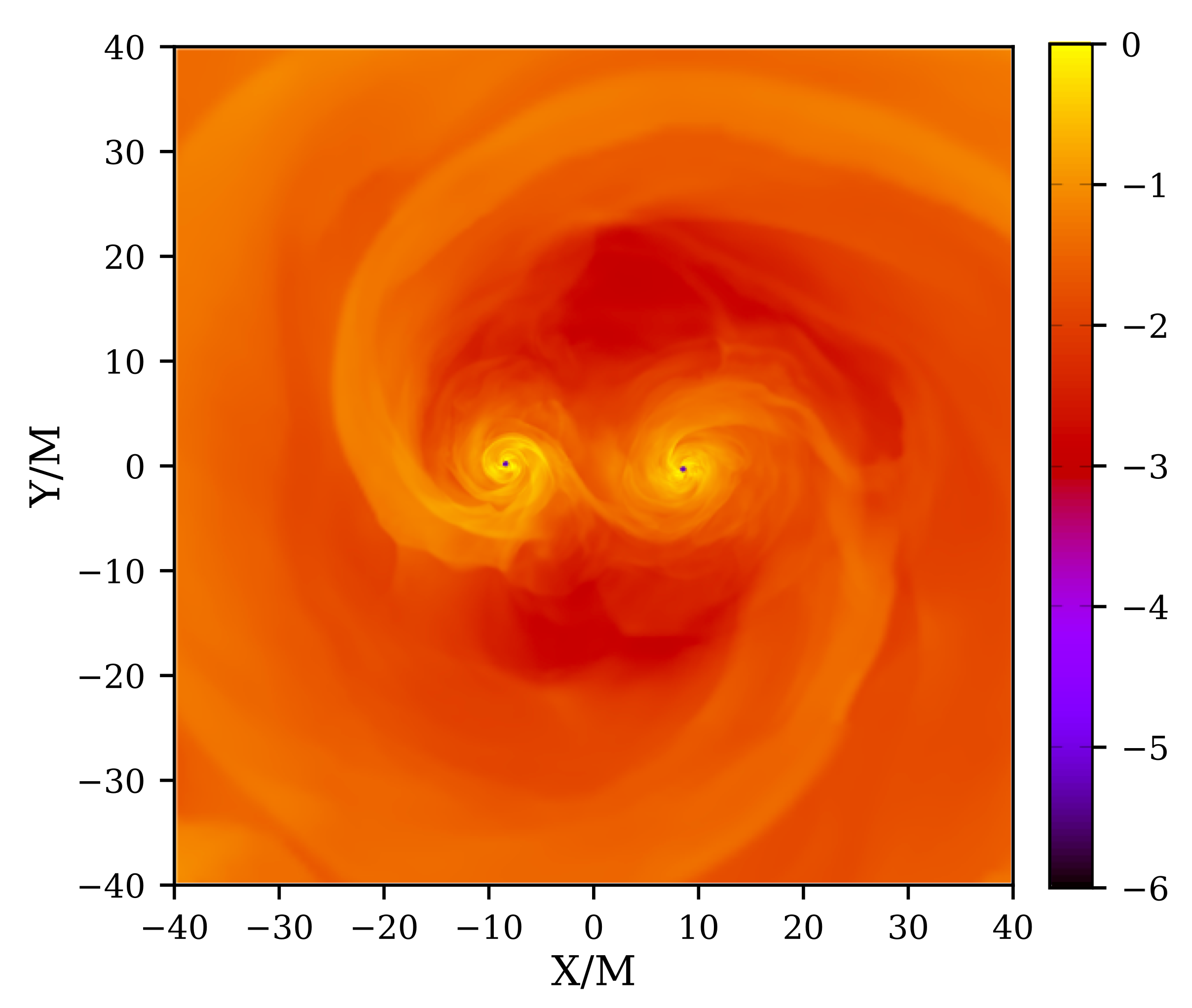}
  \caption{Equatorial rest mass density of the $\chi_{++}$ model illustrating the minidisk structures around each black hole after $\sim13$ binary orbits. 
    \label{fig:density}}
    \end{figure}

Fig. ~\ref{fig:density} shows the rest mass density in the equatorial
plane in the $\chi_{++}$ model, which exhibits clear persistent
minidisks around each of the black holes.  Fig.~\ref{fig:sigma} shows
the disk parameters $\Sigma(\varpi)$, and $\alpha(\varpi)$ which are
the surface density profile of the minidisks, and the effective
viscosity of the minidisks, all of which have been azimuthally and
time averaged (over $\sim 7$ full binary orbits) to create a 1D radial
profile centered on the minidisk. The parameters reported are
calculated for one of the black holes in the $\chi_{++}$ model and the
positive spin black hole in the $\chi_{+-}$ model.

The surface density profile shows the clear presence of a minidisk
with a peak surface density around $3.2M$ in the $\chi_{++}$ minidisk,
and closer to $4M$ in the $\chi_{+-}$ minidisk. The effective
viscosity parameter $\alpha$ has a value of about 0.02 in most of the
$\chi_{++}$ disk, and a bit higher in the $\chi_{+-}$ disk, and then
shoots up to much larger values in the inner region inside of the ISCO
as expected. Outside of about $6M$, which is close to the approximate
Hill sphere radius at these binary separations, $\alpha$ becomes
negative, indicating that the flow in this region is not disk-like,
hence it is not a well-defined quantity. Ref. \cite{Gold:2014a} also
found similar small values of $\alpha$ in the circumbinary disk in
cases where they did not use a cooling prescription. The time and
azimuthally averaged scale height divided by the radius in the
minidisks is approximately constant with $H(\varpi)/\varpi\simeq
0.28$.  Thus, minidisks are puffy structures.

\begin{figure}
  \centering
  \includegraphics[width=0.47\textwidth]{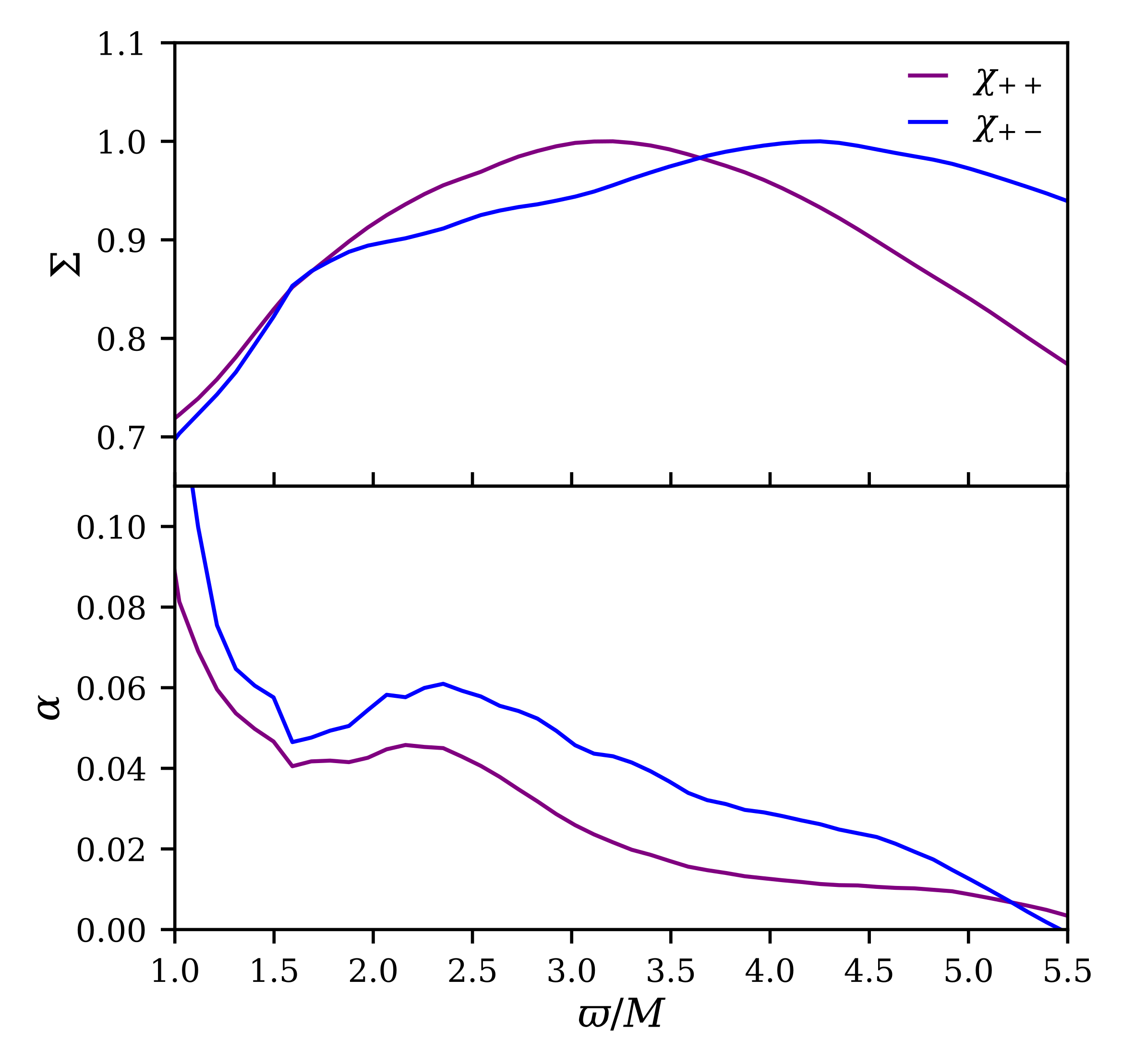}
  \caption{Azimuthally and time averaged minidisk structure parameters for the $\chi_{++}$ and $\chi_{+-}$ models. \textit{Top:} surface density profile, \textit{bottom:} effective viscosity parameter. Here $\varpi$ indicates the cylindrical radius centered on the black hole. The Hill sphere radius is at about $7M$,  and the ISCO radius for the positive spin black hole is $r_{ISCO} = 1.58M$. (Note that here we use $M = m_1 +m_2$ the total mass of the system, not the mass of the individual black hole). 
    \label{fig:sigma}}
    \end{figure}

\comment{
\begin{figure}
  \centering
  \includegraphics[width=0.47\textwidth]{../paper2_plots/upup/plot_sigma_all.png}
    \includegraphics[width=0.47\textwidth]{../paper2_plots/updown/plot_sigma_all.png}
  \caption{surface density
    \label{fig:sigma}}
    \end{figure}
    
 \begin{figure}
  \centering
  \includegraphics[width=0.47\textwidth]{../paper2_plots/upup/plot_h_all.png}
    \includegraphics[width=0.47\textwidth]{../paper2_plots/updown/plot_h_all.png}
  \caption{height
    \label{fig:h}}
    \end{figure}
    
 \begin{figure}
  \centering
  \includegraphics[width=0.47\textwidth]{../paper2_plots/upup/plot_alpha_all.png}
    \includegraphics[width=0.47\textwidth]{../paper2_plots/updown/plot_alpha_all.png}
  \caption{alpha
    \label{fig:alpha}}
    \end{figure}
    
\subsection{Outflows and Jets}
\label{jets}

\begin{figure}
  \centering
  \includegraphics[width=0.53\textwidth]{../paper2_plots/upup/mag.png}
  \caption{magnetization
    \label{fig:sigma}}
    \end{figure}
}

\subsection{Variability}
\label{variability}

In this section we explore the periodic nature of the accretion rates
and mass contained within the Hill spheres, as well as the impact of
the existence of persistent minidisks on these properties. In this
work we build from our previous finding that at our initial separation
of $20M$, all black holes in our studies with $\chi=+0.75$ or $\chi=0$
form persistent minidisks, while all black holes in our studies with
$\chi=-0.75$ did not and mass was quickly accreted.

We investigate periodicities through computation of the power spectral
density (PSD = $\mid FFT \mid ^2$) of these signals across each of our
models. We normalize the frequencies to the average binary orbital
frequency, $f_{orb},$ for each model.

As noted in our previous work, the time-averaged accretion rate is
affected by the black hole spins, with the negatively spinning black
holes exhibiting a higher average accretion rate than positively
spinning black holes. This is consistent with the negative spin black
holes being unable to form minidisks, and thus material from the
accretion streams plunges into the BHs directly. Additionally, we
found that the average rest-mass within the Hill spheres was much
greater for the positively spinning black holes (prograde spin) than
the negatively spinning (retrograde spin), again consistent with the
formation of persistent minidisks in the positively spinning cases,
and hence longer inflow time from the minidisks as opposed to the case
where the tidal streams plunge into the black holes.  In addition to
the difference in time-averaged quantities across our models, we also
observe clear periodicities in the accretion rates as well as in the
mass within the Hill spheres, which will be the focus of our
discussion here.

\begin{figure*}
  \centering
  
   \includegraphics[width=\textwidth]{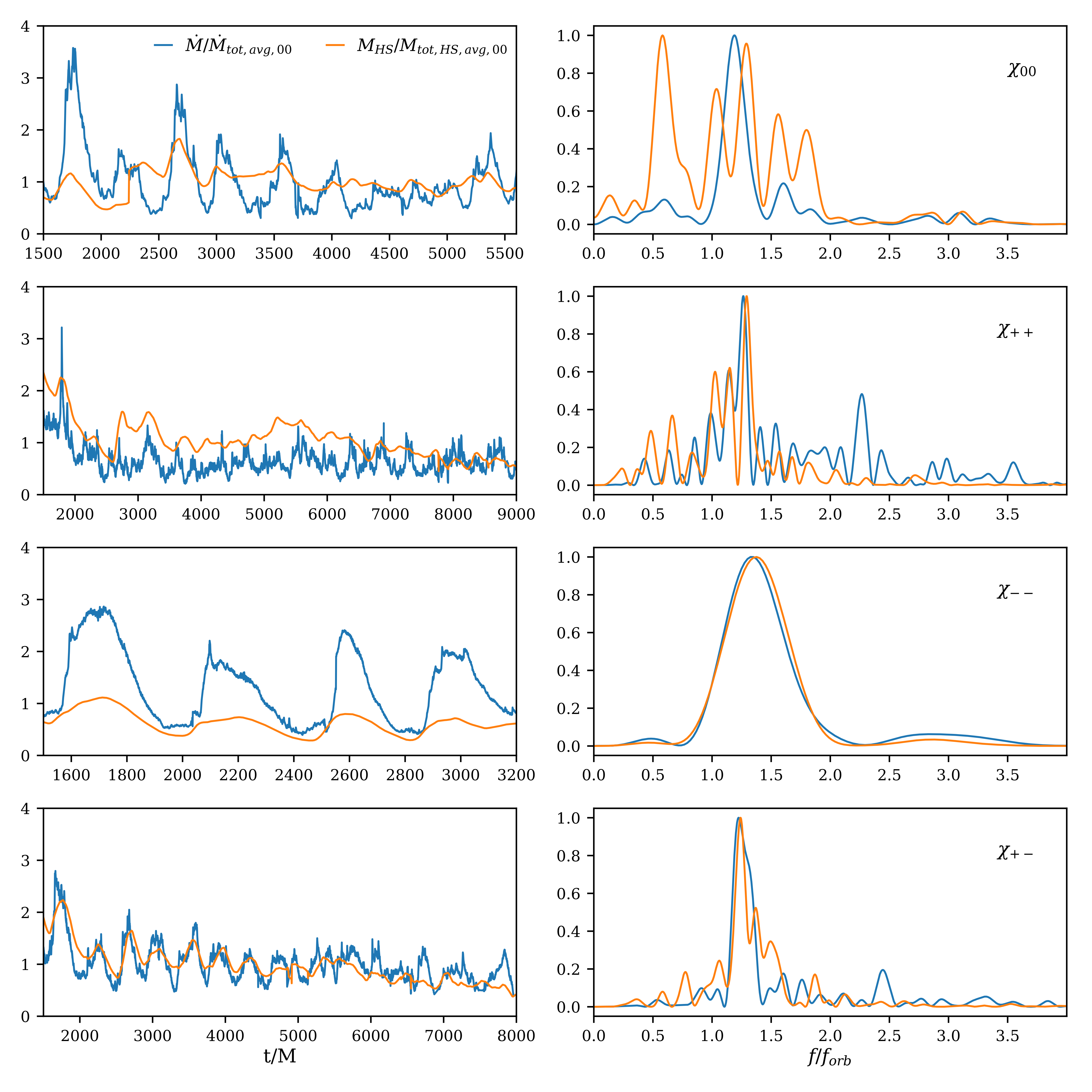}

  \caption{\textit{Left}: Total accretion rate ($\dot M$) (blue
    curves) and total mass contained within the Hill spheres
    ($M_{HS}$) (orange curves) as a function of time for each of our
    models. From top to bottom we show the $\chi_{00}, \chi_{++},
    \chi_{--}$ and $\chi_{+-}$ models. We plot the combined total of
    of these quantities for both black holes in all models. Both
    accretion rate and mass within the Hill spheres exhibit periodic
    fluctuations with time that are mostly in phase with one
    another. The reported quantities are normalized by their
    respective average values in the $\chi_{00}$ model to provide a
    consistent normalization across all models. \textit{Right:} PSD of
    the accretion rate and mass within the Hill spheres shown in the
    corresponding left panel normalized to the peak value of the PSD
    in each case.
    \label{fig:acc}}
    \end{figure*}

\begin{figure}
  \centering
  \includegraphics[width=0.48\textwidth]{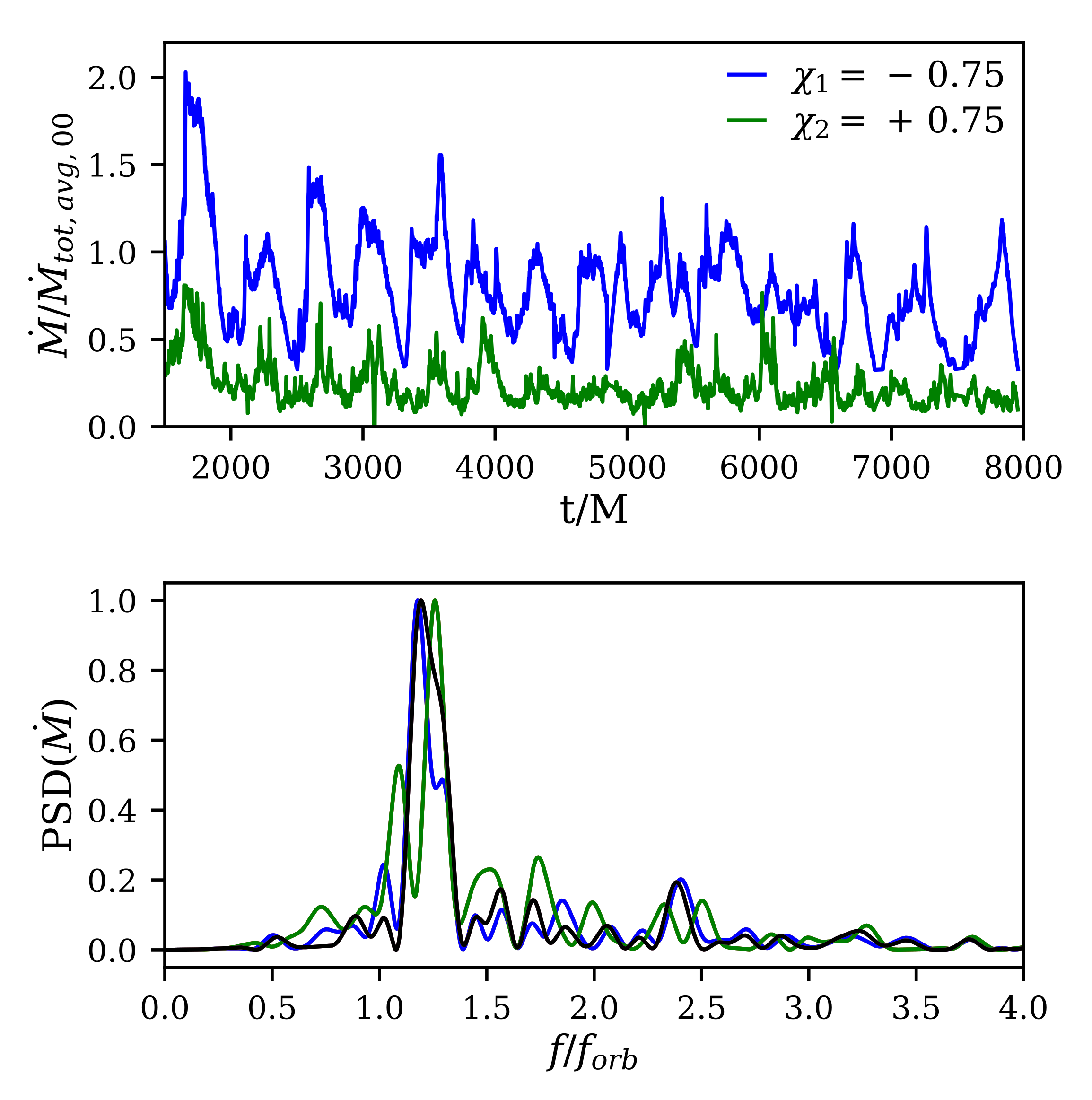}
  \caption{\textit{Top:} Accretion rates onto the individual
    black holes in the $\chi_{+-}$ model, illustrating the effect of
    the minidisk around the positively spinning black hole causing a
    significantly dampened amplitude in the modulation of the
    accretion rate compared to the negatively spinning black hole
    where no minidisk is present. \textit{Bottom:} PSD of the above
    accretion rates, with the black curve representing the PSD of the
    total accretion rate. Each PSD is normalized to its own peak value. 
  \label{fig:udacc}}
    \end{figure}

\begin{figure}
  \centering
  \includegraphics[width=0.48\textwidth]{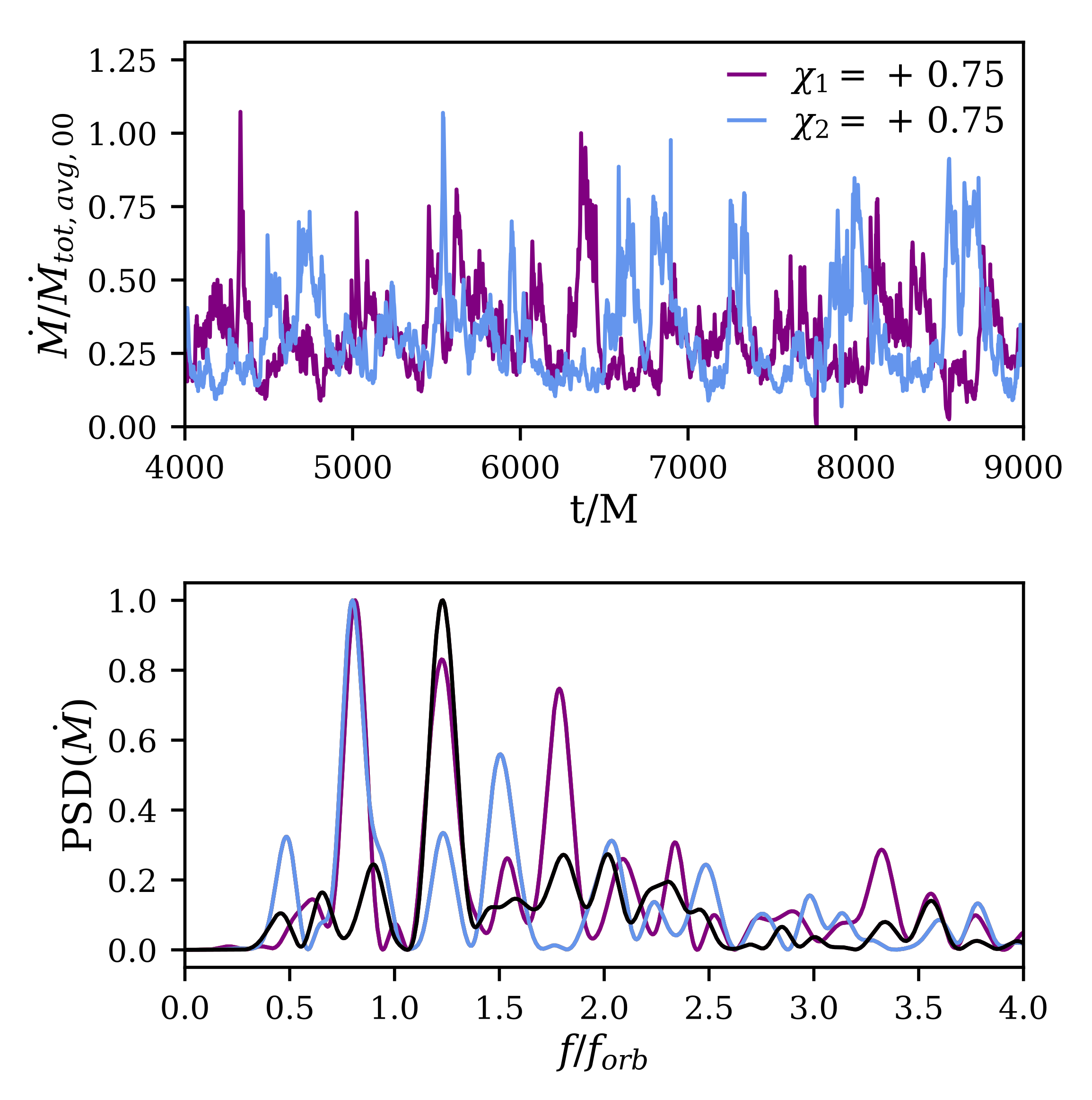}
  \caption{\textit{Top:} Accretion rates onto the individual
    black holes in the $\chi_{++}$ model.  \textit{Bottom:} PSD of the
    above accretion rates, with the black curve representing the PSD
    of the total accretion rate. Each PSD is normalized to its own peak value. 
  \label{fig:uuacc}}
    \end{figure}

\begin{figure}
  \centering
  \includegraphics[width=0.48\textwidth]{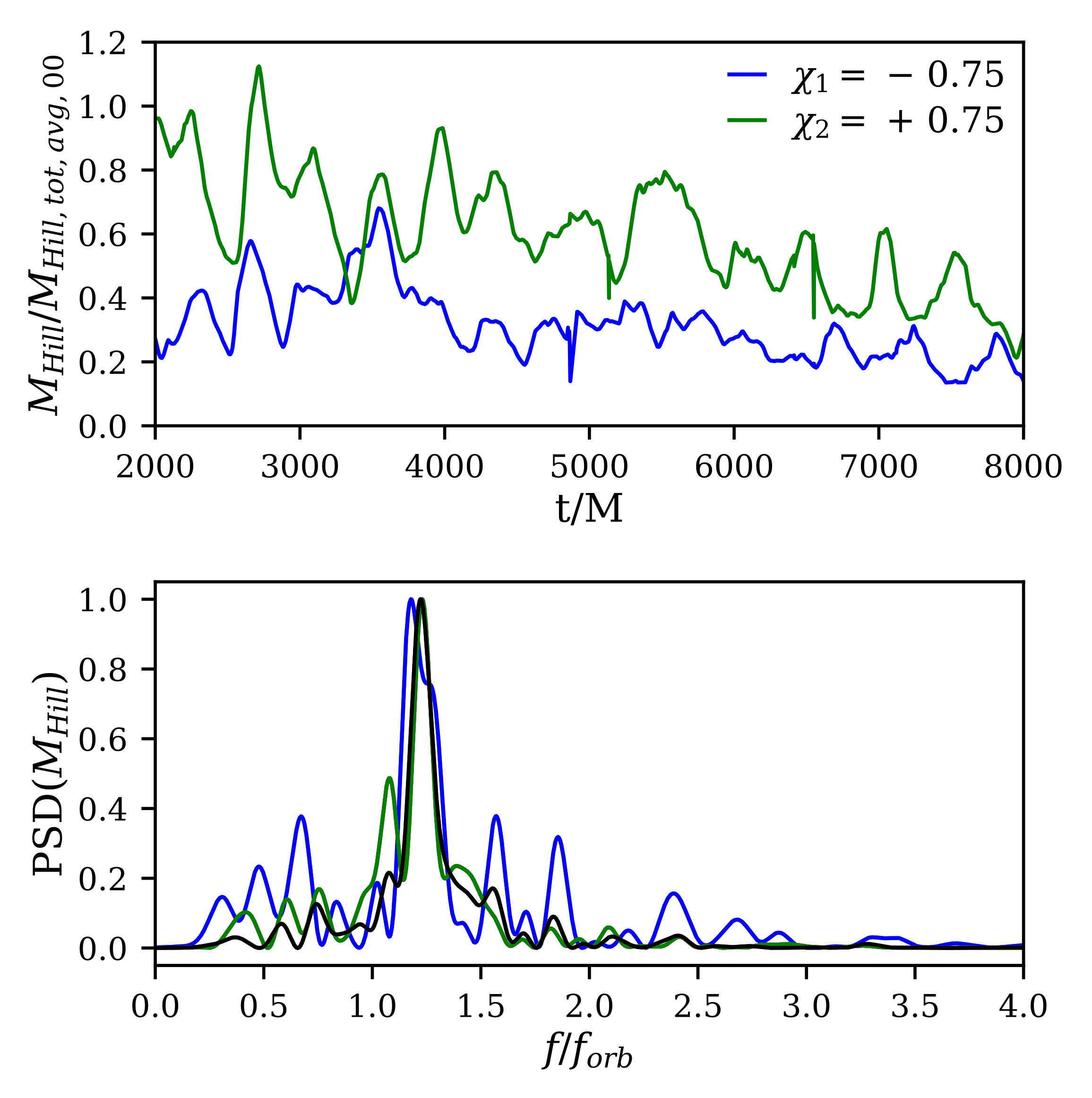}
  \caption{\textit{Top:} Masses within the Hill spheres of
    the individual black holes in the $\chi_{+-}$ model.
    \textit{Bottom:} PSD of the above masses, with the black curve
    representing the PSD of the total mass contained in both Hill
    spheres. Each PSD is normalized to its own peak value. 
  \label{fig:udmass}}
    \end{figure}  
    
\begin{figure}
  \centering
  \includegraphics[width=0.48\textwidth]{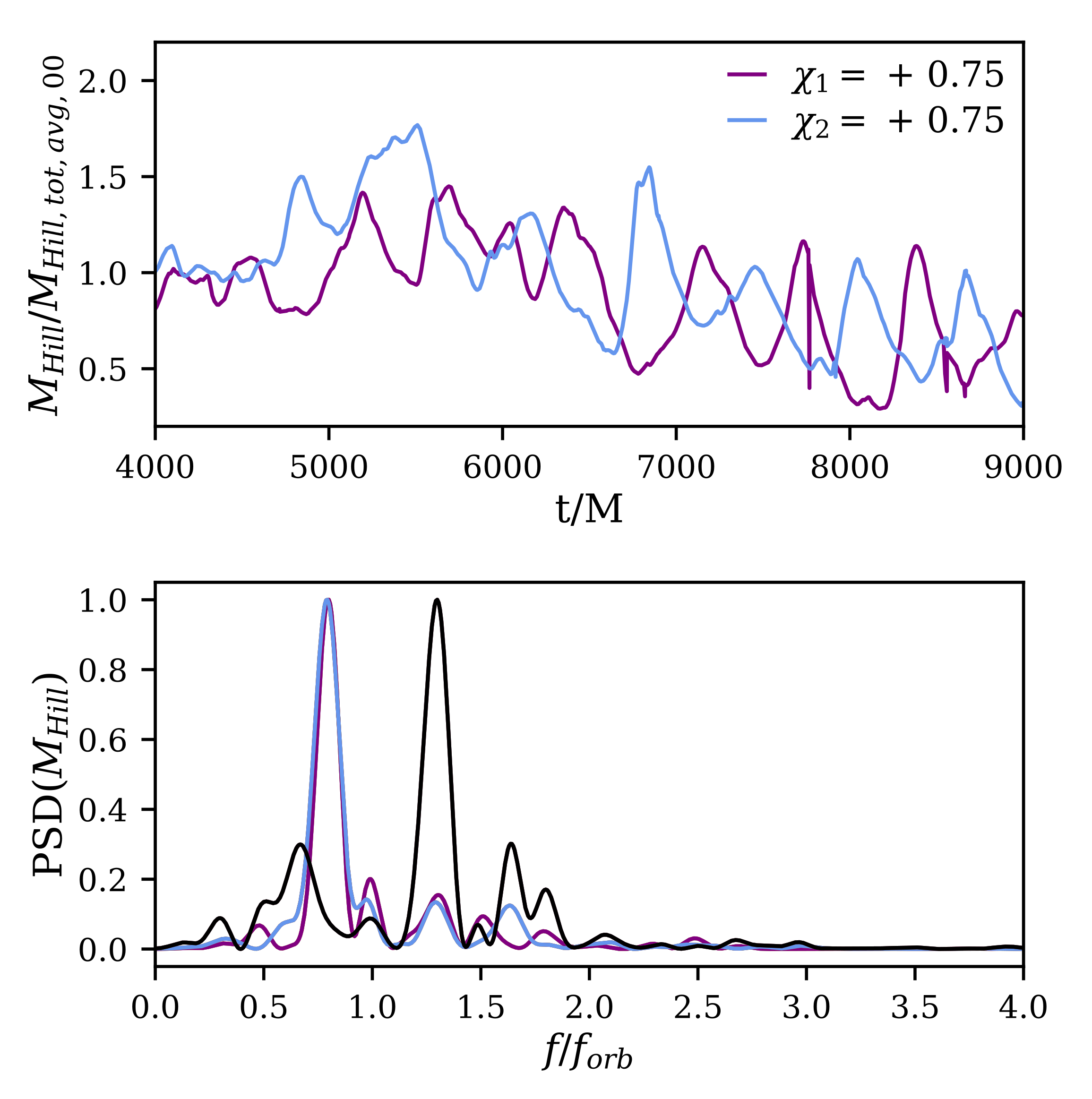}
  \caption{\textit{Top:} Masses within the Hill spheres of
    the individual black holes in the $\chi_{++}$ model.
    \textit{Bottom:} PSD of the above masses, with the black curve
    representing the PSD of the total mass contained in both Hill
    spheres. Each PSD is normalized to its own peak value. 
  \label{fig:uumass}}
    \end{figure}

As shown in the right panels of Fig. \ref{fig:acc}, we find that all
of our models exhibit a definitive peak or peaks in the PSD of the
accretion rate, indicating a periodicity at a particular frequency in
the accretion. We find the most prominent peak in the PSD of the
accretion rate in all models falls between $\sim1.2-1.4 f_{orb}$
consistent with previous
studies~\cite{Noble:2012,Shi:2015,Bowen:2018,Bowen:2019,Combi:2022}. The
PSD of the mass contained within the Hill spheres shows strong peaks
that closely match those in the accretion rates in each model. The
$\chi_{00}$ case also exhibits several other strong peaks in the PSD
of the mass contained within the Hill spheres that are not present in
the accretion rate, the strongest of which occur at $\sim f_{orb}$ and
$\sim 0.5 f_{orb}$.

We also find that the modulations in the accretion rates and the mass
within the Hill spheres are predominantly in phase, as shown in the
left panels of Fig. \ref{fig:acc}. This correlation between the
frequencies and phase of the modulations in the accretion and the mass
within the Hill spheres is not surprising, as cycles of increased
rest-mass within the Hill spheres leads directly to an increase in the
rest-mass accreted onto the black holes. This is in agreement with the
findings of \cite{Combi:2022}, which describes the time-dependence of
the minidisk mass as a smoothed version of that of the accretion
rates.

We further investigate the relative strength of the modulation by
comparing the amplitudes of the variability in the accretion rates and
mass within the Hill spheres. This demonstrates that the modulation is
strongest in the negative spinning model, and becomes weaker with
increasingly positive spin. This can be seen easily in the amplitudes
of the variability in the left panel of Fig. \ref{fig:acc} by
contrasting the $\chi_{00}$, $\chi_{++}$, and $\chi_{--}$ models (top
three panels). This is also clearly demonstrated by comparing the
accretion rates onto the individual black holes in the $\chi_{+-}$
case as seen in Fig.~\ref{fig:udacc}. These demonstrate that {\it the
  presence of a persistent minidisk not only slows the accretion rate,
  but also suppresses the strength of the quasi-periodic modulation in
  the accretion rate}.

We quantify this by calculating the ``average deviation'', which
measures the average distance of the data set from the mean accretion
rate, as well as calculating the root-mean-square (RMS).  These values
are reported in Table~\ref{tab:rms}.  Note that since the accretion
rates are normalized by the average accretion rate in the $\chi_{00}$,
this allows us to compare the amplitudes of the variation in each
model. Also note that these normalized accretion rates are of order
unity (see left panels of Fig \ref{fig:acc}), hence the numbers
reported in Table~\ref{tab:rms} can be viewed as crude
percentage-level fluctuations. We find a monotonic decrease in both
the average deviation and the RMS with increasing spin values in the
$\chi_{--}$, $\chi_{00}$, and $\chi_{++}$ models, with a factor of
$\sim3$ decrease in both the RMS and the average deviation comparing
the $\chi_{++}$ model to the $\chi_{--}$ model.  In last two rows of
the table we also report the RMS and average deviation from the mean
in the $\chi_{+-}$ case for each black hole separately, where it is
clear that the prograde spin black hole exhibits much smaller
fluctuations than the retrograde spin black hole.

These results demonstrate that the existence of persistent minidisks
is strongly correlated with a decreased strength of the variability in
the accretion rate onto the black holes. Given that higher spins, thus
existence of persistent minidisks, are also associated with lower
accretion rates (seen clearly in Fig.~\ref{fig:udacc}), these results
suggest that for larger minidisks, the dampening of the fluctuations
arises because the inflow time from the minidisks begins to become
comparable to the minidisk feeding timescale through the circumbinary
accretion streams. 

The above conclusion may impact the interpretation of observed
periodicities in quasars as arising by the modulation of the accretion
rate onto the binary. Binaries at larger separations and therefore
larger Hill spheres and potentially much larger minidisks may have
periodicities dampened even more, and it is unclear whether they will
be able to exhibit any observable periodicities. Therefore,
quasi-periodic behavior in observed lightcurves arising from
modulations in the accretion rate may not be smoking-gun evidence for
the existence of a binary. More work is necessary to shed light on
this important effect, which will be the subject of a future paper.

\begin{table}
\centering
\caption{Root-mean square (RMS) and average deviation of the accretion rates in each of our models. All values are normalized by the same value of the average accretion rate in the $\chi_{00}$ model. Note that the values reported for the $\chi_{++}$, $\chi_{00}$, and $\chi_{--}$ are for the total accretion rate onto both black holes, while the values of the $\chi_{+-}$ model are reported for the accretion rates onto the individual black holes. }
\begin{tabular}{ ccc }
  \hline
  \hline 
 Model & RMS & Average Deviation \\
 \hline
 \hline 
 $\chi_{++}$ & 0.21 &  0.17 \\
 $\chi_{00}$ & 0.49 & 0.38 \\
 $\chi_{--}$ & 0.65 & 0.57 \\
  $\chi_{+-, +}$ & 0.10 & 0.078 \\
   $\chi_{+-,-}$ & 0.28 & 0.22 \\
  \hline
\end{tabular}
  \label{tab:rms} 
  \end{table}
 
We further examine the nature of the periodic behavior by
analyzing the accretion rates and masses within the Hill spheres for
the individual black holes for the $\chi_{+-}$ and $\chi_{++}$
models. These can be found in Figs. \ref{fig:udacc}, \ref{fig:uuacc},
\ref{fig:udmass}, \ref{fig:uumass}, along with the PSDs of these
quantities.

We find that the individual accretion rates onto the two black holes
in the $\chi_{+-}$ model appear to have the modulation of their
accretion rates to be mostly in phase, and the PSD of the individual
and total accretion rates all exhibit the same peak frequency, which
can be seen in Fig. \ref{fig:udacc}. A very similar pattern is seen in
the mass within the Hill spheres of the individual black holes in the
$\chi_{+-}$ model, where the variability is roughly in phase, and the
PSD exhibits the same peak frequency for each black hole as well as
the total mass contained in both Hill spheres, which can be seen in
Fig. \ref{fig:udmass}. Conversely, the individual accretion rates in
the $\chi_{++}$ model appear mostly out of phase. The PSD of the
individual accretion rates exhibit peak frequencies at $\sim0.75
f_{orb}$, close to half of the peak frequency of the total accretion
rate (at which the individual accretion rates also exhibit peaks in
the PSD albeit with reduced power), which can be seen in
Fig. \ref{fig:uuacc}. The same trend is present in the mass within the
individual Hill spheres in the $\chi_{++}$ model, with the modulation
out of phase, and the strongest peak in the individual PSDs at
$\sim0.75 f_{orb}$, and the strongest peak in the total Hill spheres
mass PSD at $\sim1.2 f_{orb}$, which can be seen in
Fig. \ref{fig:uumass}.

\subsection{Outflows and Jets}
\label{jets}

\begin{figure}
 \centering
  \includegraphics[width=0.45\textwidth]{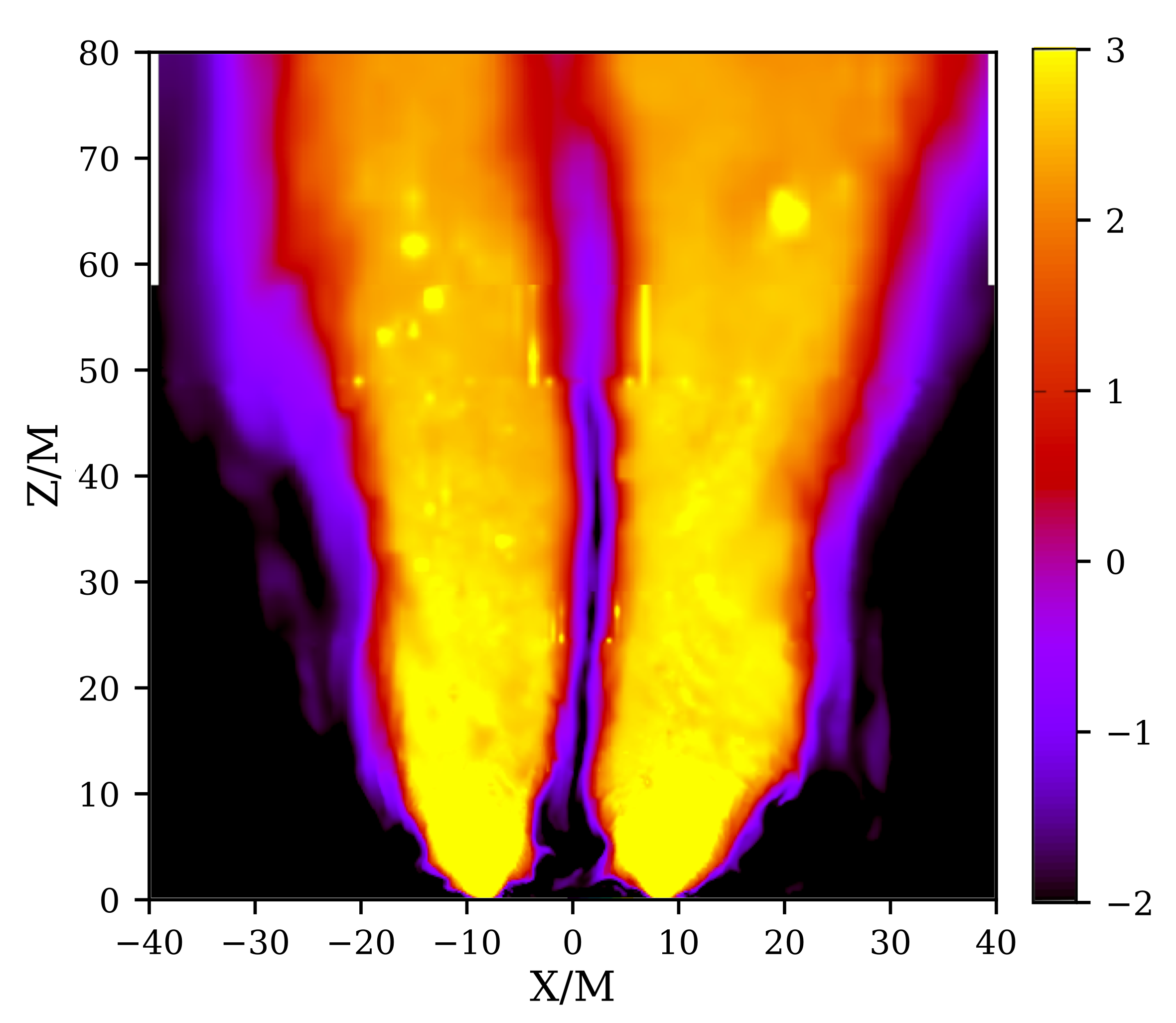} 
    \includegraphics[width=0.45\textwidth]{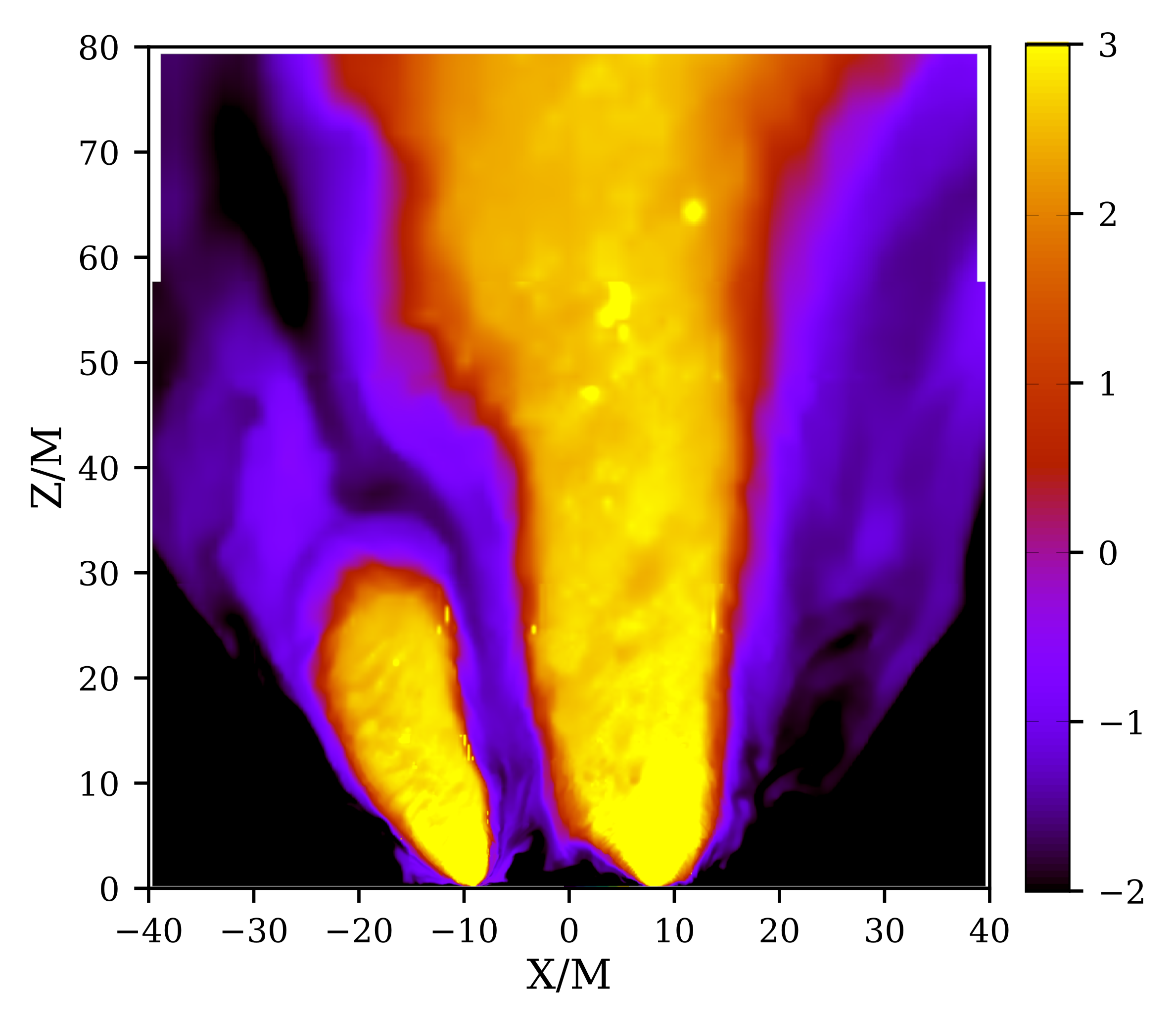}
      \includegraphics[width=0.45\textwidth]{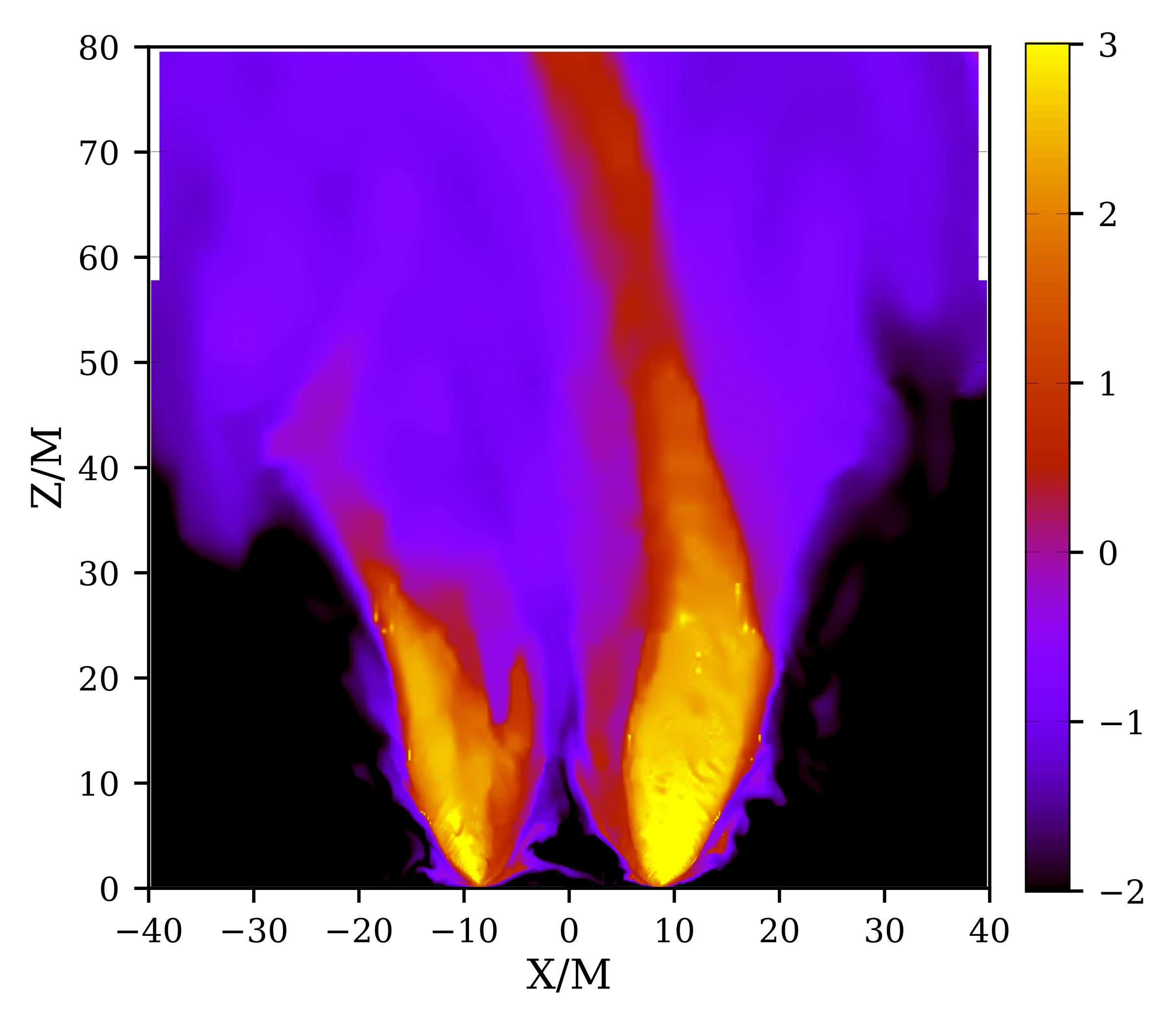}
      \caption{Meridional slice of magnetization $\log(b^2/2 \rho_0)$
        for the $\chi_{++}$ (top panel), $\chi_{+-}$ (middle panel),
        $\chi_{00}$ (bottom panel). The snapshot is chosen at a time
        where both black holes are located along the X-axis in their
        orbits, with the positive spin black hole on the right
        (positive X-position) and the negative spin black hole on the
        left (negative X-position) for the $\chi_{+-}$ case. The snapshots are taken after 13, 10, and 12 		binary orbits respectively for the $\chi_{++}$, $\chi_{+-}$, and $\chi_{00}$ models.
          \label{fig:mag}}
    \end{figure}

We observe collimated, highly magnetized outflows from the polar
regions of the black holes. In Fig. ~\ref{fig:mag} we show meridional
slices of the magnetic-to-rest-mass energy density $b^2/2 \rho_0$ of
the $\chi_{++}$, $\chi_{+-}$, and $\chi_{00}$ models.  The large
values of $b^2/2 \rho_0$ found in the outflows indicate that these
regions are magnetically dominated and the jets are magnetically
powered. These regions are nearly force-free, which is a requirement
for the Blandford-Znajek (BZ) mechanism~\cite{Blandford:1977}. We find
significantly higher values of $b^2/2 \rho_0$ in our spinning cases
than previous non-spinning fully relativistic binary accretion studies
(see e.g. \cite{Gold:2013,Gold:2014b}). The $\chi_{++}$ model shows
both black holes generating strongly magnetized outflow regions, while
in the $\chi_{+-}$ model, the positive spin black hole has a
magnetized outflow region which dominates over that of the negative
spin black hole. This is in agreement with \cite{Tchekhovskoy:2012},
which found that single black holes with spins prograde with the
accretion disk have more powerful jets than black holes with
retrograde spins. The $\chi_{00}$ model shows significantly less
strongly magnetized outflows, indicating that the spin of the black
holes plays the predominant role in the magnetization strength.  The
significantly enhanced magnetization and outgoing Poynting luminosity
in the $\chi_{++}$ and $\chi_{+-}$ models compared to the $\chi_{00}$
model, demonstrates that the traditional spin-induced BZ effect
dominates over the ``orbital'' BZ
effect~\cite{Palenzuela:2010xn}. This also demonstrates that the boost
in Poynting luminosities reported post-merger in simulations of
initially non-spinning black
holes~\cite{Gold:2014a,Gold:2014b,Khan:2018} is due to the fact that
accretion onto the remnant black hole proceeds onto a spinning black
hole.

\begin{figure}
  \centering
  \includegraphics[width=0.47\textwidth]{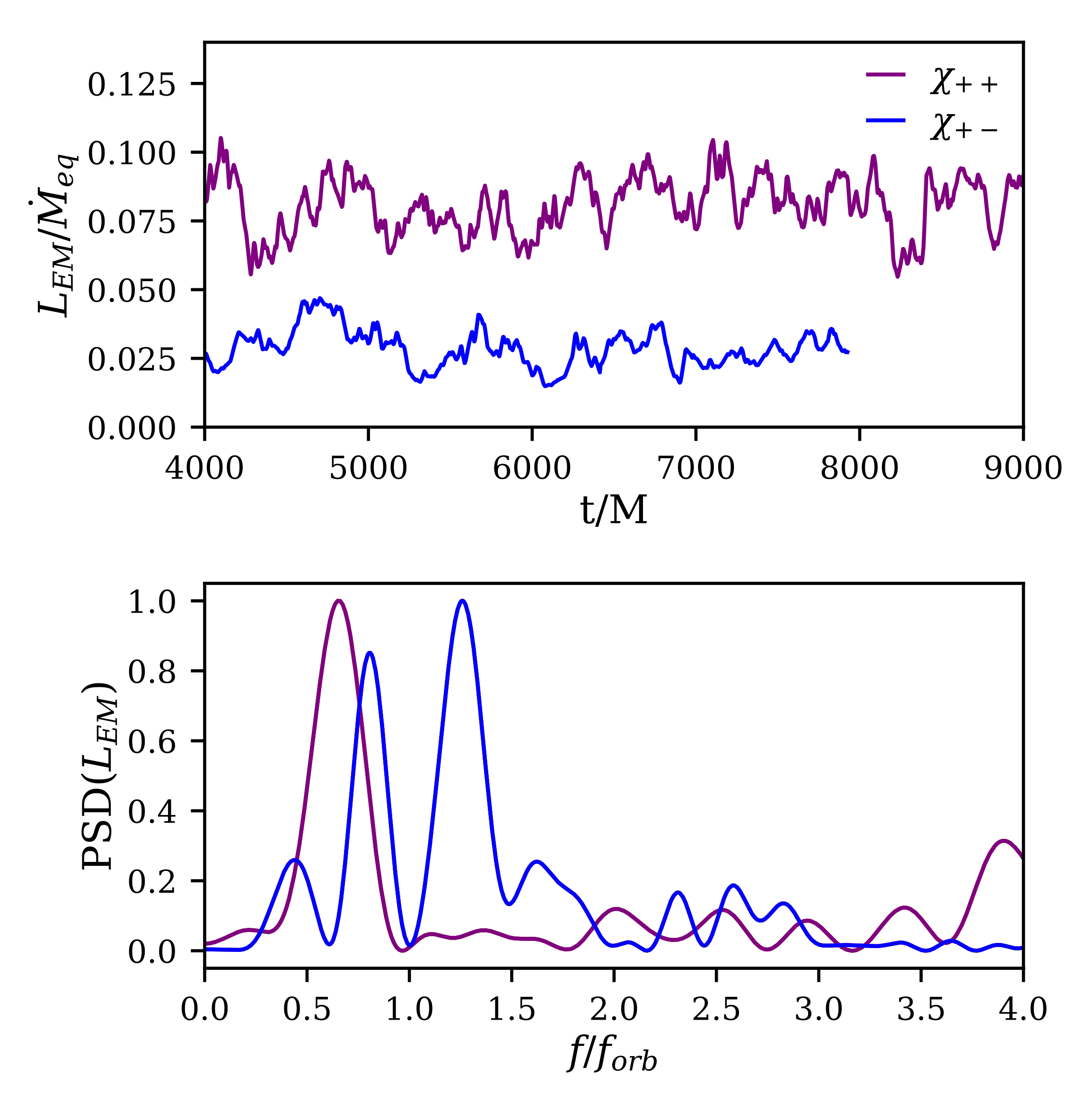}
  \caption{\textit{Top:} Poynting luminosity, $L_{EM}$, as a function of time for the $\chi_{++}$ and $\chi_{+-}$ models. \textit{Bottom:} PSD of the Poynting luminosities for each model normalized to the peak value of the PSD in each case. 
    \label{fig:Lem}}
    \end{figure}

We also examine the Poynting luminosities (calculated far away from
the binary-disk system and considered only after the quantity settles
following the initial burst) and their PSDs focusing on the
$\chi_{+-}$ and $\chi_{++}$ cases, which can be seen in
Fig. \ref{fig:Lem}.  Both the $\chi_{+-}$ and $\chi_{++}$ models show
a clear peak in the PSD of the Poynting luminosity, though they do not
have their peaks at the same frequency across the models like was seen
in the accretion rates. In the PSD of the $\chi_{+-}$ model, we see a
peak at $\sim 1.2 f_{orb}$, approximately equal to the peak seen in
the PSD of the accretion rate (either total or onto the individual
black holes). However, the PSD of the $\chi_{++}$ model has its peak
at $\sim 0.6 f_{orb}$, about half of the peak seen in the PSD of the
total accretion rate. However, this is approximately equal to the
dominant peaks seen in the individual accretion rates. It is unclear
as to why this is the case, but in our case we compute the
luminosities on the surface of a distant sphere that encompasses the
entire binary+circumbinary disk system. In the $\chi_{+-}$ case the
Poynting luminosity is dominated by the prograde spin BH hence the
prograde spin accretion rate variability is reflected in the jet.
Given that that the accretion time series onto the individual black
holes in the $\chi_{++}$ are not in phase, it is not unreasonable to
expect that the Poynting luminosity exhibits a periodicity at the
accretion rate periodicity onto the individual BHs, which exhibit peak
observed variability at $\sim 1/2$ of the main frequency of the total
accretion rate.

\section{Conclusions}
\label{conclusions}

We have continued our work from \cite{Paschalidis:2021} to extend the
evolution of our GRMHD simulations in full GR and further our analysis
of minidisks and black hole spin in accreting SMBH binaries.

We present details of the structure of minidisks though 1D orbit and
time averaged profiles of the surface density, scale height, and
effective viscosity that may be helpful to inform the parameters used
in Newtonian studies in which binaries approach the relativistic regime.

We examine the periodic behaviors of the mass of the minidisks, the
accretion rates, and the Poynting luminosities, and make comparisons
across our four models. We find a clear peak in the PSD of both the
total mass of the minidisks and total accretion rates corresponding to
$\sim 1.2-1.4 f_{orb}$, which is consistent with previous
simulations. We find this peak to be present in all of our models,
regardless of spin or presence of minidisks. We find a peak at roughly
the same frequency in the Poynting luminosity of the $\chi_{+-}$
model, and at roughly half that frequency in the $\chi_{++}$ model,
which is correlated with the modulation frequency of the accretion
rates onto the individual black holes in the system.

While the peak frequency of the periodicity is consistent across our
models, the strength of the modulation is not. We find a factor of
$\sim3$ reduction of the average deviation and the RMS variability in
the accretion rates in positively spinning black holes that exhibit
minidisks compared to the negatively spinning black holes where no
minidisks are present. This is found both in comparing the accretion
rates of the $\chi_{++}$ and $\chi_{--}$ models, and the individual
black holes in the $\chi_{+-}$ model. This indicates that the presence
of minidisks works to dampen out the strength of the periodic nature
of the accretion onto the black holes.  At larger separations, and
thus larger Hill spheres and larger minidisks, it is possible that the
variability may be dampened out even further, leading to little or no
periodicities at larger separations. In future work we will
investigate simulations at larger separations to better probe the
behavior of larger minidisks and periodic behaviors.

\section*{Acknowledgements}

This work was in part supported by NSF grants PHY-1912619 and
PHY-2145421 to the University of Arizona, as well as NSF Graduate
Research Fellowship grant DGE-1746060. Computational resources were
provided by the Extreme Science and Engineering Discovery Environment
(XSEDE) under grant No. TG-PHY190020. XSEDE is supported by the NSF
grant No. ACI-1548562. Simulations were performed on \texttt{Comet},
and \texttt{Stampede2}, which is funded by the NSF through award
ACI-1540931.

\section*{Data Availability}
 
The data underlying this article will be shared on 
reasonable request to the corresponding author.




\bibliographystyle{mnras}
\bibliography{ref} 





\bsp	
\label{lastpage}
\end{document}